\documentclass[aps,10pt,twocolumn,twoside,showpacs]{revtex4}

\usepackage{graphicx}
\usepackage{color}
\usepackage{epsfig}
\usepackage{amsmath}
\usepackage{amstext}
\usepackage{amssymb}
\usepackage{bm}
\usepackage[colorlinks=true,linkcolor=blue]{hyperref}

\begin{document}
\author{Mateusz Cholascinski}
\affiliation{Institut f\"ur Theoretische Festk\"orperphysik,
Universit\"at Karlsruhe, D-76128 Karlsruhe, Germany}
\affiliation{Nonlinear Optics Division, Adam Mickiewicz University,
  61614 Poznan, Poland}

\date{\today}
\title{Appearance of topological phases in superconducting nanocircuits}

\begin{abstract}
We construct non-Abelian geometric transformations in superconducting
nanocircuits, which resemble in properties the Aharonov-Bohm phase for
an electron transported around a magnetic flux line. The effective
magnetic fields can be strongly localized, and the path is traversed
in the region where the energy separation between the states involved
is at maximum, so that the adiabaticity condition is weakened. In
particular, we present a scheme of topological charge pumping. 
\end{abstract}


\maketitle

When an electron is transported in magnetic field around a closed loop, it
acquires a phase equal to the magnetic flux through the surface
spanned by the electron path. This phenomenon has been known as the
Aharonov-Bohm effect for nearly half of a century \cite{AB}. In the
original paper the magnetic field forms a flux line, and
electrons move in the region of zero field. \textcite{berry}
considered this phenomenon to be an early example of the geometric
phases, which he described in a more general context. For each quantum
system undergoing adiabatic cyclic evolution of its parameters we can
find phase shifts acquired by its energy eigenstates. 
Apart from the dynamical factor, there is a contribution which
depends only on the geometry of the path traversed in the parameter
space (the parameters are time dependent, as they are varied
throughout the process, but the explicit time dependence does not
enter the expression for the Berry phase, which makes it {\em
  geometric} in nature). Berry gives also a formula for this phase in
the form of an integral of  
an effective magnetic field (we will refer to this field as the
``Berry field'') over a surface spanned by the path. This makes
the similarity between geometric phases in arbitrary quantum system,
and the Aharonov-Bohm 
scenario even closer: one can think of the Berry field as analogous
to the real magnetic field; the parameter space corresponds then to
the real position space. The Aharonov-Bohm effect in the original setting
is, however, easier to observe than the Berry phase in general. The
first reason is that in the former case there is 
no adiabaticity condition constraining the electron velocity, while in the
latter the rate of the parameters' variation should be much smaller than
the inverse energy difference between the states involved. The second
is that for strongly localized field any variation in the
path of the electron does not affect the phase at all as long as the
path encloses the flux line (this phase is thus {\em
  topological}). For the Berry phase, depending on the system, 
we encounter effective fields usually smoothly varying with the
parameters, or even uniform, particularly for a spin-$1/2$ system, the phase
is the flux of a monopole, or, in other words, the area spanned by the
traversed loop on the unit sphere (see e.g.
\cite{berry}). Fluctuations of externally  
controlled parameters first of all lead to dephasing of dynamical origin, but
also smear the path, which affects the visibility of the effect even further. 

Here we show that geometric phases can appear in quantum systems which are
robust in the sense explained above: the effective magnetic field
obtained from the Berry formula (the Berry field) is not necessarily a
monopole (uniform in space) field, but can be strongly
localized near the 
points of the lowest energy spacing between the states involved, and
thus the geometric phases generated in the process can be topological. The
field is suppressed in the region where the energy spacing is at
maximum. This region is thus the most robust, as the fluctuations of
the parameters do not change the geometric phases. Since the energy
spacing is large in this region, the adiabaticity condition is weakened,
and the path traversed relatively fast should give strong effects. 

We find here the simplest, two-dimensional non-Abelian phases in the
system considered in Ref.~\cite{chola}, but for a different range of
the system parameters. This changes the picture
substantially, as the topological properties of the phases become now
evident (geometric 
phases in superconducting nanocircuits have 
been studied also in \cite{falci,faoro}; in the earlier
settings, however, the phases have not been proven to be
topological). In our system the 
operational subspace is a  
twofold degenerate ground state, so that the dynamical contribution is
just an overall phase factor, and the actual transformations are of purely
geometric origin. Since the fields are localized, these transformations
are topological in nature. Furthermore, we find a way to control the
localization of the Berry field, as well as
of the phase acquired while going around the region of enhanced
field. The position of the peaks can be found from the parameter
dependence of the spectrum;
the degree of the localization reflects the deviation of the imaginary
part of the Hamiltonian from zero. In the limit 
of completely real Hamiltonian, the Berry field is zero except for
singular points, and the geometric phase can have only discreet
values $\{-\pi , 0, \pi \}$. 

\begin{figure}[t] 
\centerline{\resizebox{0.35\textheight}{!}{\rotatebox{0}
{\includegraphics{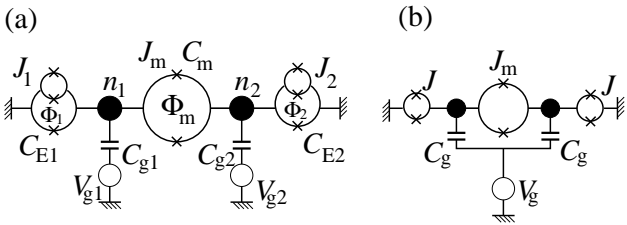}}}}
\caption{(a) The system composed of two ``charge qubits,'' where the state
  is encoded in the number of Cooper pairs on the superconducting
  islands (black nodes), {\it i.e.} $n_1$, $n_2$. The voltages and
  magnetic fluxes can be controlled in a cyclic way in order to
  perform non-Abelian geometric transformations. (b) A simplified setup
  for topological charge pumping.}
\label{system}
\end{figure}

The system we consider (see Fig.~\ref{system}) consists of two charge
qubits \cite{makhlin} coupled to each other via a dc SQUID
(superconducting quantum 
interference device). The electrostatic energy
of the system, including all the capacitances in the system, depends
on the number of the Cooper-pairs on each island [denoted by $(n_1,
n_2)$], and equals $E_{n1, n2} = E_{c1} (n_{g1} - n_1)^2 +
E_{c2}(n_{g2}-n_2)^2 + E_m (n_{g1}-n_1)(n_{g2}-n_2)$. with $E_{c1} =
4e^2 C_{\Sigma 2}/2(C_{\Sigma 1}C_{\Sigma 2} - C_m^2)$, and similarly
for $E_{c2}$. Here $C_{\Sigma 1(2)}$ is the sum of all capacitances
connected to the first (second) island. The inter-qubit electrostatic coupling 
$E_m = 4e^2 C_m/ (C_{\Sigma 1} C_{\Sigma 2} - C_m^2)$. The Josephson coupling  
$J^{(1)} = \sqrt{(J^{(1)}_{d}-J^{(1)}_{u})^2 +
  4J^{(1)}_{u}J^{(1)}_{d}\cos^2(\pi \Phi _{1})} \exp [- i \psi 
(\Phi _{1})]$ (and similarly for $J^{(2)}$), where $\tan \psi (\Phi_1
) = (J^{(1)}_{d}-J^{(1)}_{u})/(J^{(1)}_{d} + J^{(1)}_{u}) \tan \pi
\Phi _1$ \cite{Tinkham}. Here the couplings $J_d$, and $J_u$ correspond
to the lower 
and upper junctions in the dc SQUIDs, which connect the qubits to their
reservoirs. The upper junctions $J_u^{(1)}$, and $J_u^{(2)}$ are here
replaced with another dc SQUID
loops, which enables control over the symmetry of the big dc SQUIDs (the
meaning of this technique is explained in more details in
Ref.~\cite{chola}). The important fact is that the
Josephson coupling is real for $J_u = J_d$, and complex otherwise. 
Finally $J_m = J_m^{(0)} \cos \pi
\Phi _m$ is the Josephson energy of the middle SQUID. The magnetic
fluxes are here normalized to $\Phi _0 = h c/ 2 e$, the
superconducting flux quantum.  
The Hamiltonian of the system reads
\begin{eqnarray}
  \nonumber
  H &=& \sum_{n_1,n_2}^{}E_{n_1,n_2} |n_1, n_2 \rangle \langle n_1,n_2| 
  - {1\over 2} \left(J_1  |n_1, n_2 \rangle \langle
    n_1+1,n_2|\right. \\
  \nonumber
  &+& J_2
  |n_1,n_2 \rangle \langle n_1,n_2+1| + J_m |n_1+1, n_2 \rangle
  \langle n_1,n_2+1| \\ && \left. + h.c.\right). 
  \label{eq:qua}
\end{eqnarray}
For the Josephson couplings switched off ($J_1=J_2=J_m=0$), the energy
eigenstates are the charge states with their corresponding electrostatic
energies. For the particular point $n_{g1} = n_{g2} = 1/2$, the states
$|\bar{0} \rangle \equiv |01 \rangle  $ and $|\bar{1} \rangle \equiv
|10 \rangle $ span twofold degenerate ground state. We construct
the geometric transformations by 
choosing an appropriate closed path in the parameter space, which begins
(and ends) at this point (and we refer to it as to the starting
point). If the gate voltages $n_{g1}$ and $n_{g2}$
are close to $1/2$, only four states are relevant ($n_{1,2} = 0,1$),
and the calculation of the geometric transformations corresponding to
given paths is straightforward. \\
In particular \footnote{A detailed description of these schemes  
is presented in \cite{chola}, so here we give only short overview}, if
we keep the couplings $J_1 = J_m = 0$, while $J_2$ complex and finite,
the states
$|\bar{0} \rangle $ and $|\bar{1} \rangle $ belong to two decoupled
subspaces of the Hilbert space. By varying adiabatically two parameters,
  e.g. $n_{g2}$ and $\Phi _2$ around a loop, we generate a phase shift
between those states [$\exp (i \phi _1 \sigma _z/2)$ in the basis
$\{|\bar{0} \rangle, |\bar{1} \rangle  \}$] equal to the difference of
the Berry phases acquired by the states. The 
remaining independent   
parameter $n_{g1}$ is adjusted accordingly during the variation, so
that the degeneracy in the ground state is maintained.  Similarly for
parameters 
specified by $J_1 = J_2 = J$ and $E_{01} = E_{10}$, the states $|+
\rangle = (|\bar{0} \rangle + |\bar{1} \rangle )/\sqrt{2}$ and $|-
\rangle = (|\bar{0} \rangle - |\bar{1} \rangle )/\sqrt{2}$ belong to
decoupled subspaces, and the geometric phase shift between them 
again equals to the difference of their Berry phases. In the latter
case, however, the corresponding transformation is $\exp (i \phi _2
\sigma _x/2)$, so that the two transformations generate a non-Abelian
group of (all) unitary transformations within the subspace $\{|\bar{0}
\rangle, |\bar{1} \rangle \}$. 

\begin{figure}[t] 
  \centerline{\resizebox{0.27\textheight}{!}{\rotatebox{0}
      {\includegraphics{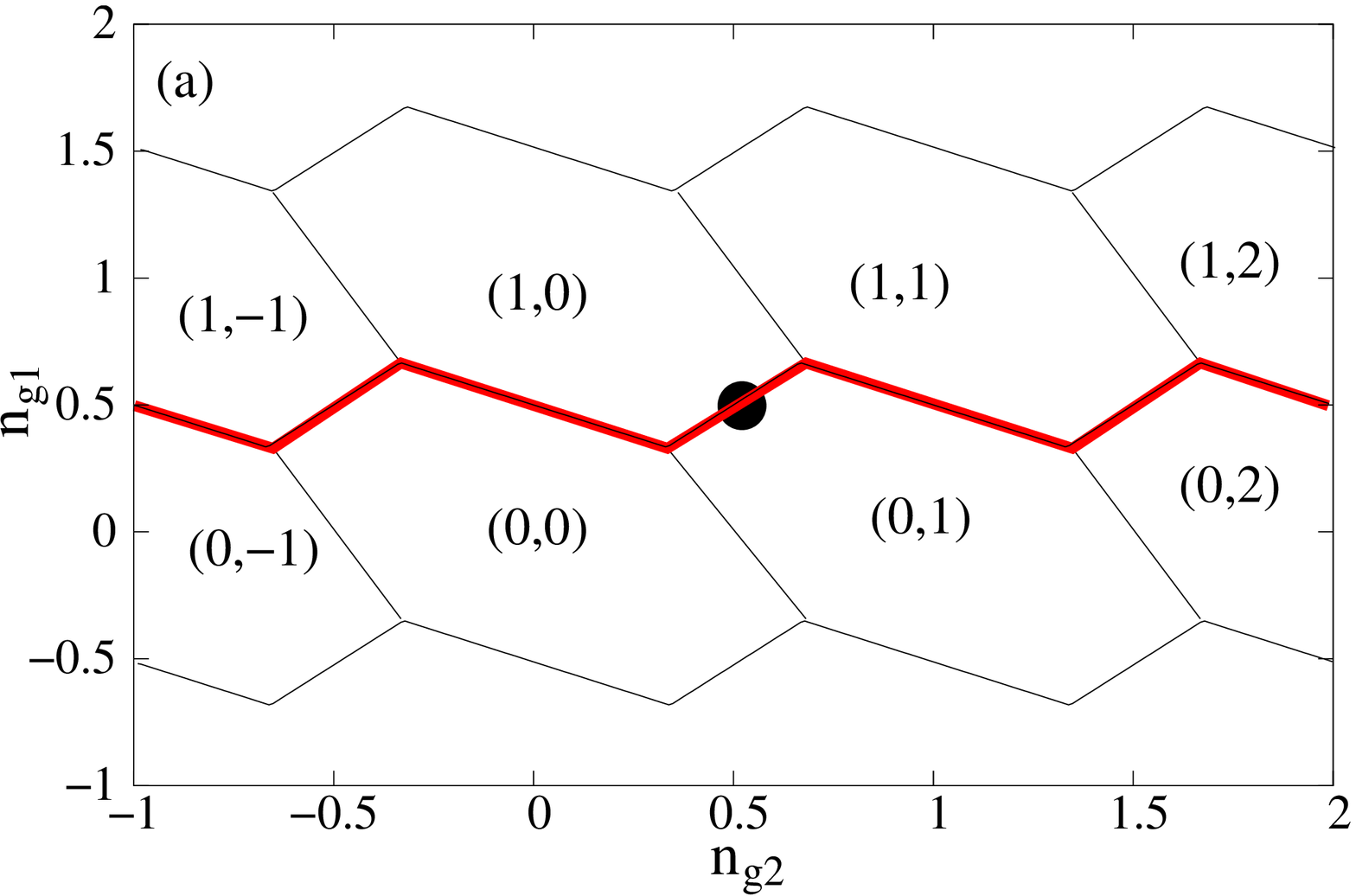}}}}
  \centerline{\resizebox{0.27\textheight}{!}{\rotatebox{0}
      {\includegraphics{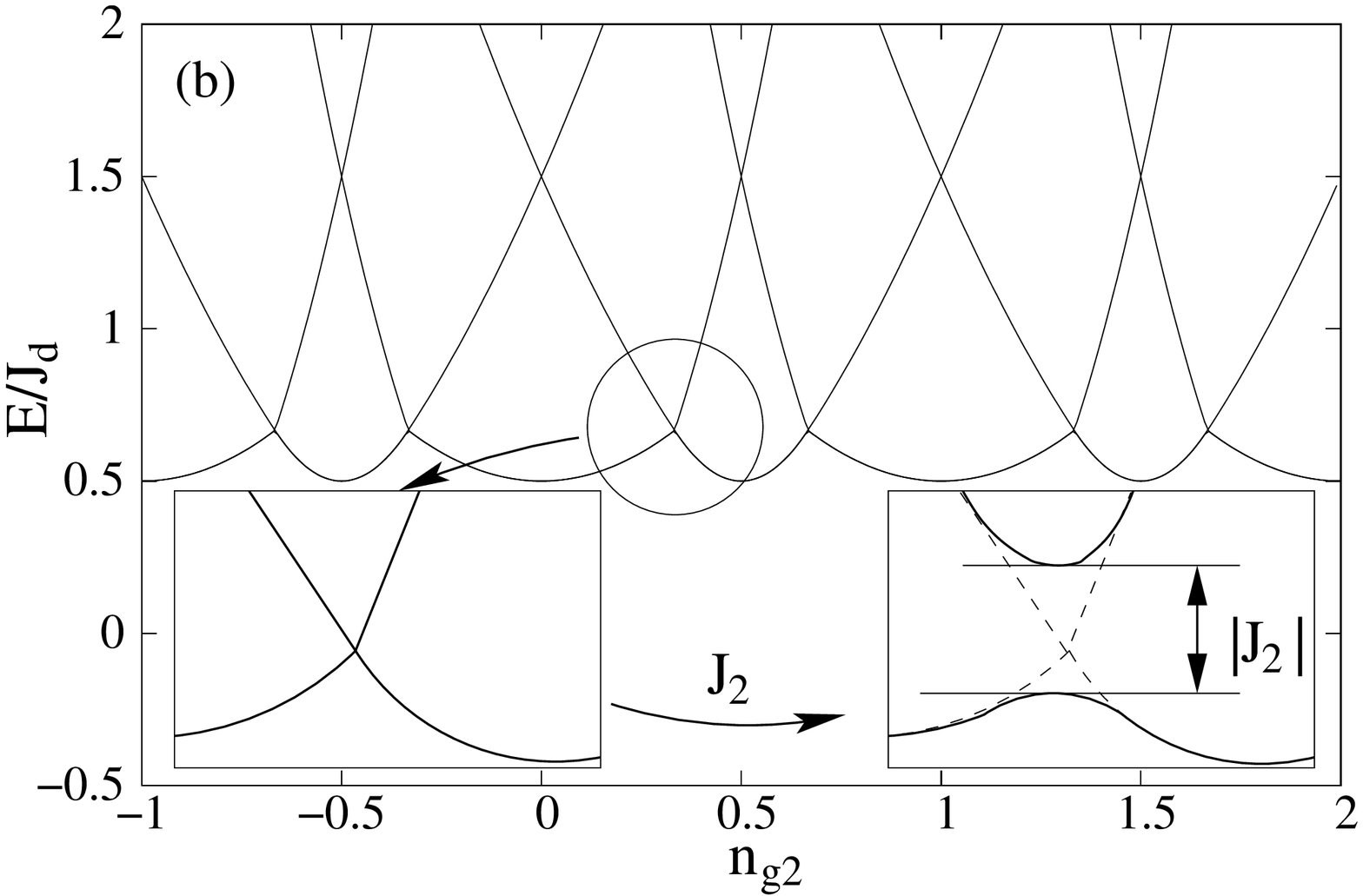}}}}
  \caption{(a) Ground state charging diagram and (b) electrostatic
      energies corresponding to the thick line in (a). The black node
      denotes hereafter the starting point. Inset: nonzero
      Josephson coupling mixes the states so that there is an avoided
      level crossing near the ``triple points'' in (a).}
  \label{honeyelect}
\end{figure}

Let us now recall the original Berry formula for the geometric
phase. The Hamiltonian is smoothly varied in time with its parameters,
$H[R(t)]$, and for each point $R(t)$ of the parameter space its eigenstates
are $|n[R(t)] \rangle $ with the energies $E_n [R(t)]$. The geometric phase
corresponding to the path $C$ in the parameter space, 
acquired by the state $|n \rangle $, equals then
\begin{equation}
   \gamma_n (C) = - \int_{S(C)} dS\cdot V_n (R),
  \label{eq:rua}
\end{equation}
where the integral is evaluated over the surface spanned by $C$ of the
effective magnetic (Berry) field
\begin{equation}
    V_n(R) = \Im \sum_{m\neq n} {\langle n | \nabla_R H |m \rangle
    \times \langle m | \nabla_R H |n \rangle \over (E_m(R) - E_n(R))^2}.
  \label{eq:sua}
\end{equation}
The energy difference in the denominator makes it possible to predict
the regions of suppressed and enhanced field. Let us consider the
spectral characteristics of our system. In Fig.\ref{honeyelect}(a) we
see exemplary diagram of the ground-state charging energies. If we
move along the borders between the cells, the ground state remains
twofold degenerate. In Fig.\ref{honeyelect}(b) the spectrum
corresponding to the thick line in (a) is plotted as the function of
$n_{g2}$. 
If we now want to perform a phase shift between the states $|\bar{0}
\rangle $ and $|\bar{1} \rangle $, we switch the coupling $J_2$ to a
finite value. The coupling will strongly mix the states near the
points of
crossing with the first excited state, as shown in the right inset of
Fig.~\ref{honeyelect}(b). However, even for finite coupling, the
minimum in the energy separation between the ground state, and the
first excited state will correspond to $n_{g2}$ of the triple
points in Fig.\ref{honeyelect}(a). For
these points, the spacing further depends on $|J_2|$, which is tuned
by the flux $\Phi _2$ and has a minimum value for $\Phi _2 = 1/2 + k$ and
maximum for $\Phi _2 = k$, $k$ being an integer. The same discussion
applies to the second transformation, 
$\exp (i \phi _2 \sigma _x/2)$. Assuming for simplicity that the design is
symmetric (characteristics of the charge qubits are identical), we
expect minimum spacing for $\Phi \equiv \Phi _1 = \Phi _2 = 1/2 +k$,
and $n_g \equiv n_{g1} = 
n_{g2}$ corresponding to the triple points ($\Phi _m$ is in this
case used to maintain the degeneracy in the ground state). On the
other hand, the
field should be strongly suppressed (and it is, as shown below) in the
region of maximum spacing between the ground state and the first
excited state. Since the aforementioned transformations are performed
by varying both, voltages and fluxes, which tune the electrostatic, and the 
Josephson energies, the optimal system should have the characteristic
Josephson,  
and electrostatic energies are of comparable magnitude. [In
Ref.\cite{chola} the transformations are constructed in the charge
regime (electrostatic energies are much bigger than the Josephson
energies) and the range of parameters considered there lies between
the minima of energy spacing, so that the Berry field seems to be
smooth and weak.]

\begin{figure}[t] 
  \centerline{\resizebox{0.37\textheight}{!}{\rotatebox{0}
      {\includegraphics{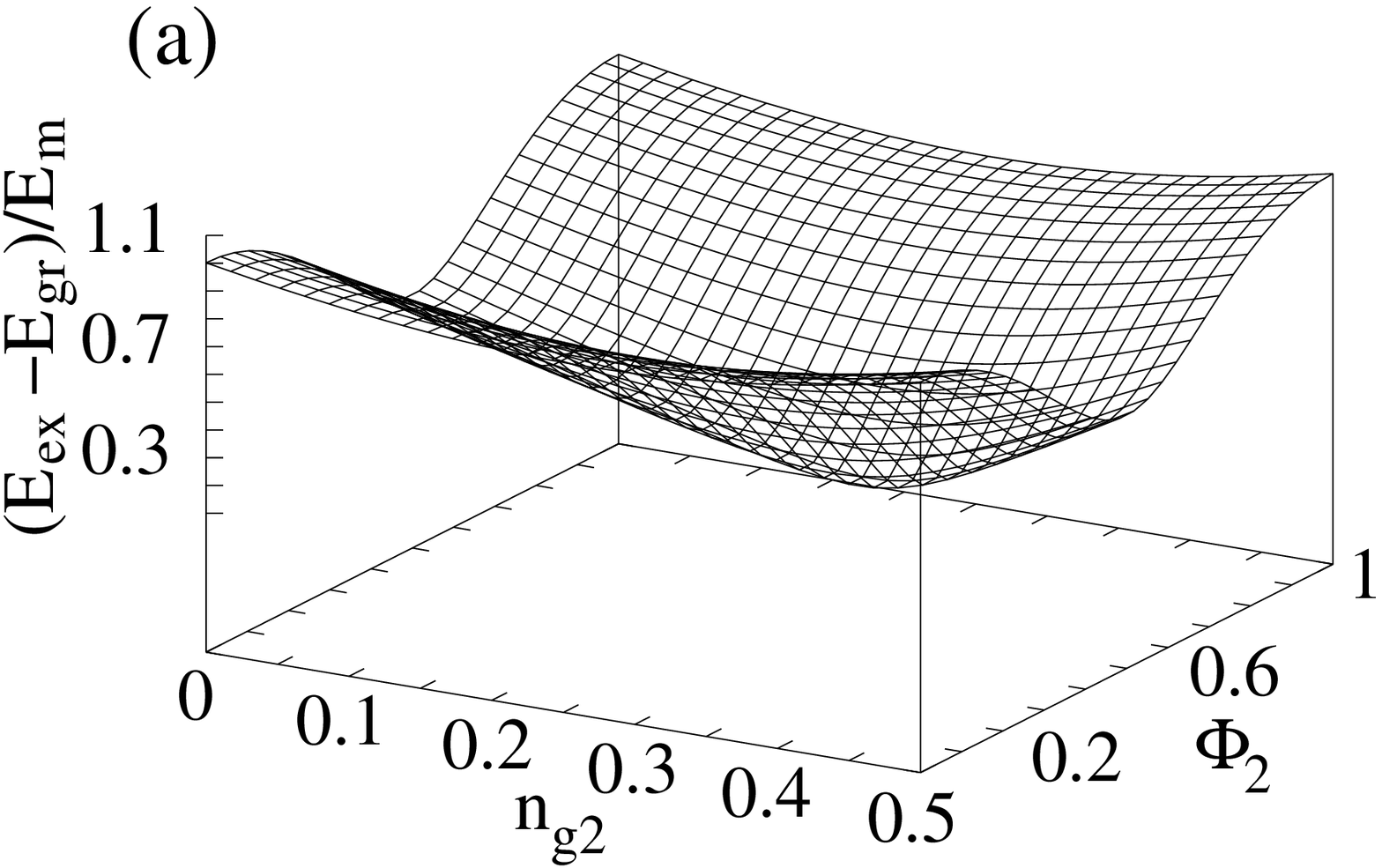}\includegraphics{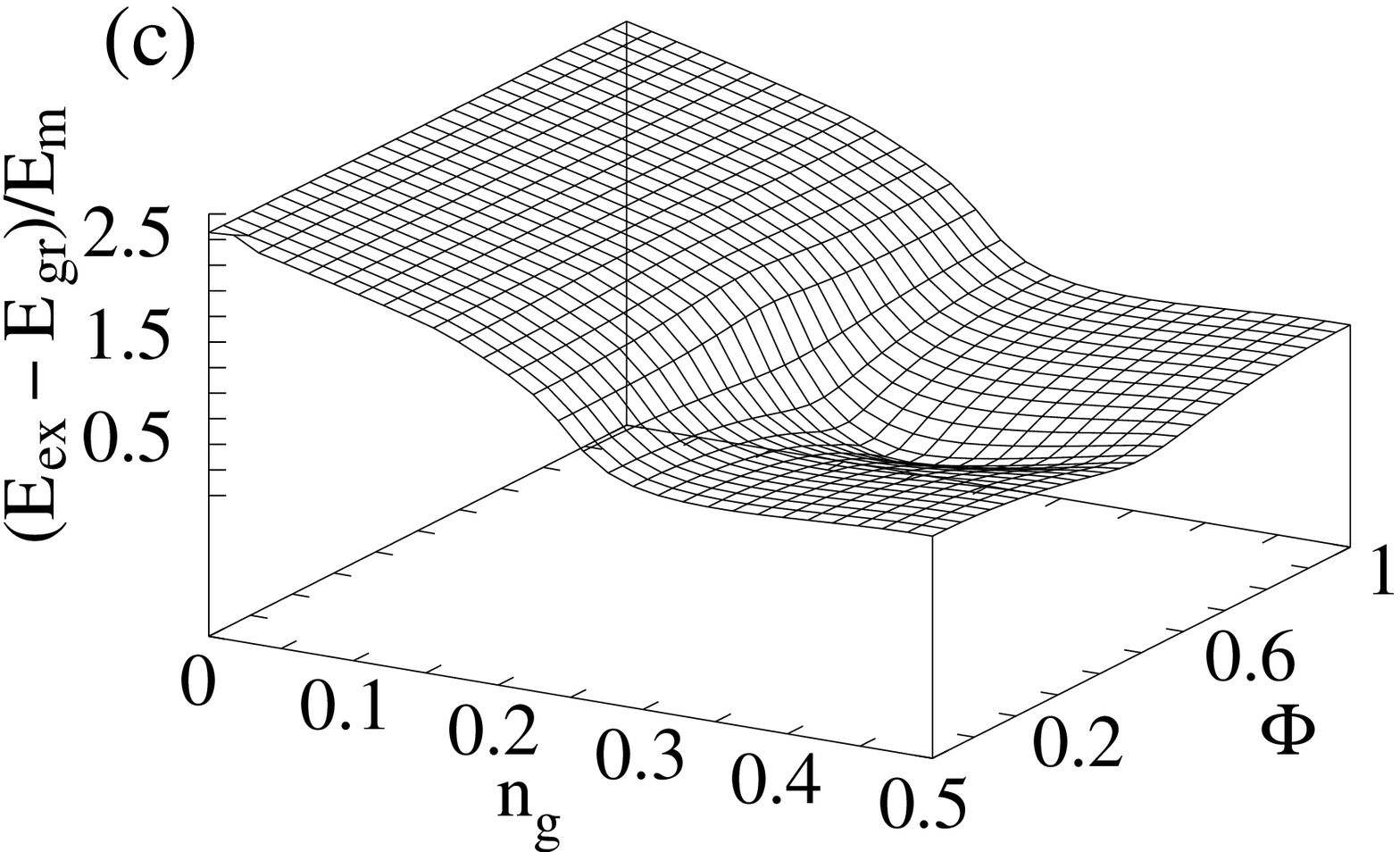}}}}
  \centerline{\resizebox{0.37\textheight}{!}{\rotatebox{0}
      {\includegraphics{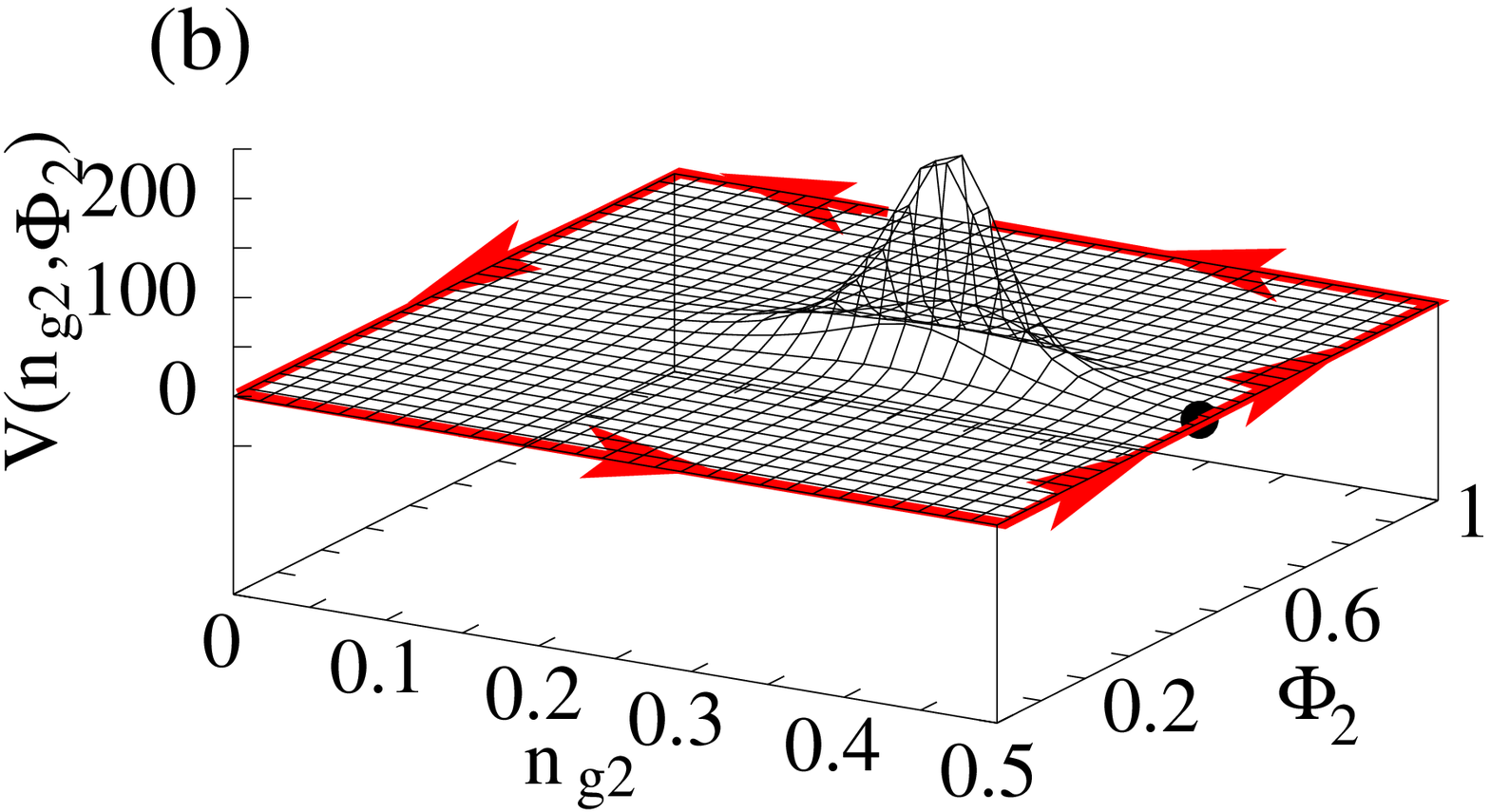}\includegraphics{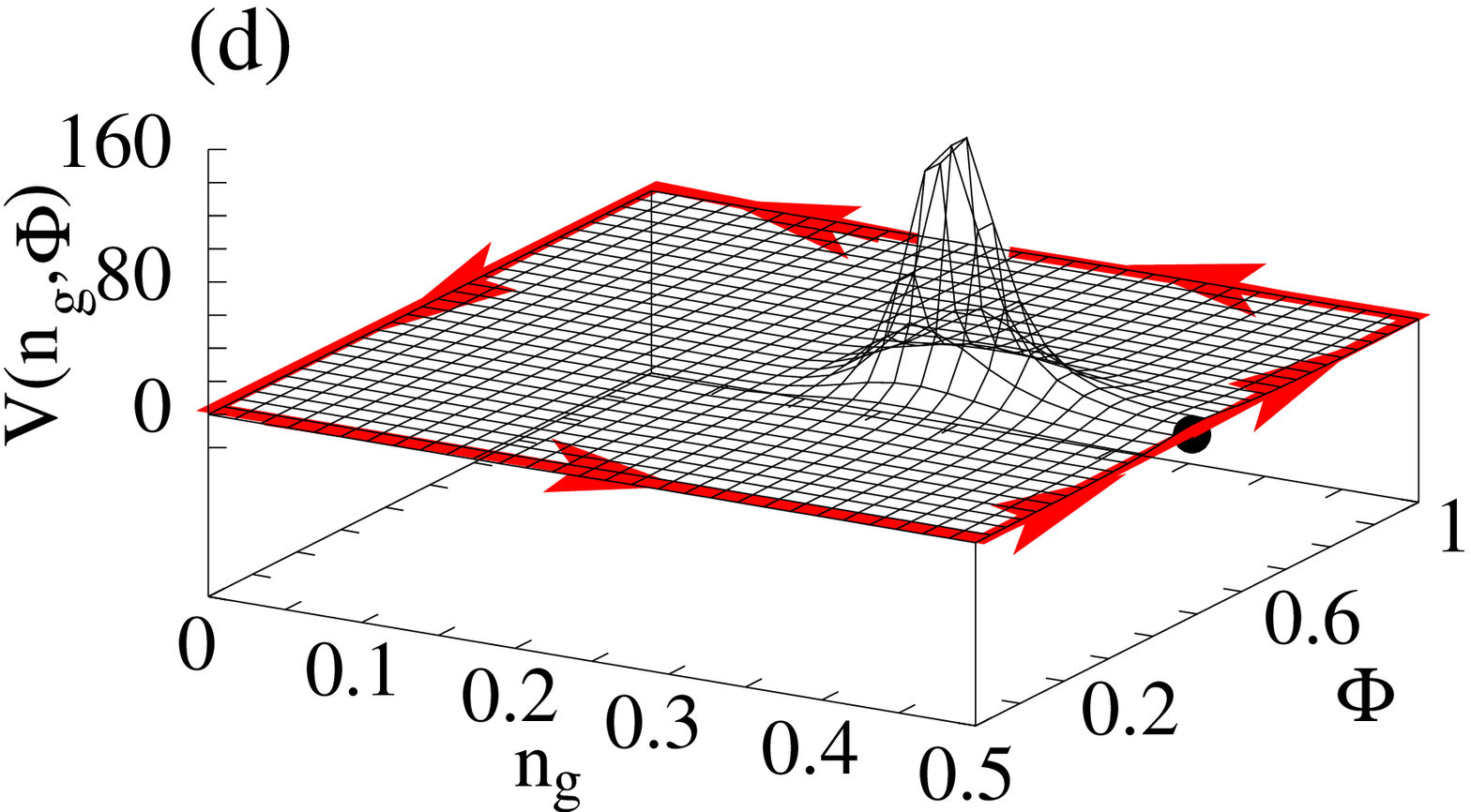}}}}
  \caption{(a) Parameter-dependent spectrum, and (b) effective
      magnetic field. The minimum in the energy spacing corresponds to
      the maximum in the field strength. Variation of the parameters
      around the peak results in a phase shift between $|\bar{0}
      \rangle $ and $|\bar{1} \rangle $. (c), (d) Similar
      characteristics corresponding to the phase shift between $|+
     \rangle $ and $|- \rangle $. The contours denote the most robust paths.}
\label{phaseflip}
\end{figure}

In Fig.\ref{phaseflip}(a) we see the difference in energy between the 
first excited and the ground state (for $J_m = J_1 = 0$, $E_{c1} =
E_{c2} = E_m$, $J_d^{(2)}/E_m = 1/2$, and $J_u^{(2)}/E_m = 2/5$), as
the function 
of $n_{g2}$ and $\Phi _2$. The gate voltage $n_{g1}$ is adjusted so
that the ground state remains degenerate. In Fig.\ref{phaseflip}(b) we see the
Berry field for the same range of parameters. The peak position
corresponds indeed to the minimum in the energy separation. The most
robust paths should be placed far from the peak, and at maximum energy
separation (so that the adiabaticity condition is weak). In our case
the optimal path is the border of the rectangle $(n_{g2}, \Phi _2) \in
[0,0.5] \times [0,1] = S$ [actually the peaks (and dips) form a lattice in the
$n_{g2}$-$\Phi _2$ plane, so that the most robust paths are borders of all
Wigner-Seitz primitive cells of the lattice].
One turn along the path produces a phase shift $\exp(i \phi _1 \sigma
_z/2)$, where
\begin{equation}
  \phi _1 = \int_S dn_{g2} d\Phi_2 V (n_{g2}, \Phi
  _2). 
  \label{eq:tua}
\end{equation}
Similarly, for $J_1 = J_2 = J$, and $n_g = n_{g1} = n_{g2}$ (for 
simplicity we assume that the qubits are identical) we vary $n_{g}$
and $\Phi  = \Phi _1 = \Phi _2$ to perform the transformation $\exp (i
\phi_2 \sigma _x/2)$ ($\Phi _m$ is tuned to maintain the degeneracy in
the ground state), where
\begin{equation}
  \phi _2 = \int_{S} dn_g d \Phi V (n_g, \Phi ).
  \label{eq:uua}
\end{equation}
The field $V(n_g, \Phi )$, for $J_d^{(1),(2)}/E_m = 1/2$ and
$J_u^{(1),(2)}/E_m = 2/5$,  is shown in Fig.\ref{phaseflip}(d), and the
spectral characteristics for the same range of parameters in
Fig.\ref{phaseflip}(c). The optimal path is here again the border of the
rectangle $(n_g, \Phi ) \in [0,0.5] \times [0,1] = S$ or of any other
Wigner-Seitz cell of the lattice. 

\begin{figure}[t] 
\centerline{\resizebox{0.38\textheight}{!}{\rotatebox{0}
{\includegraphics{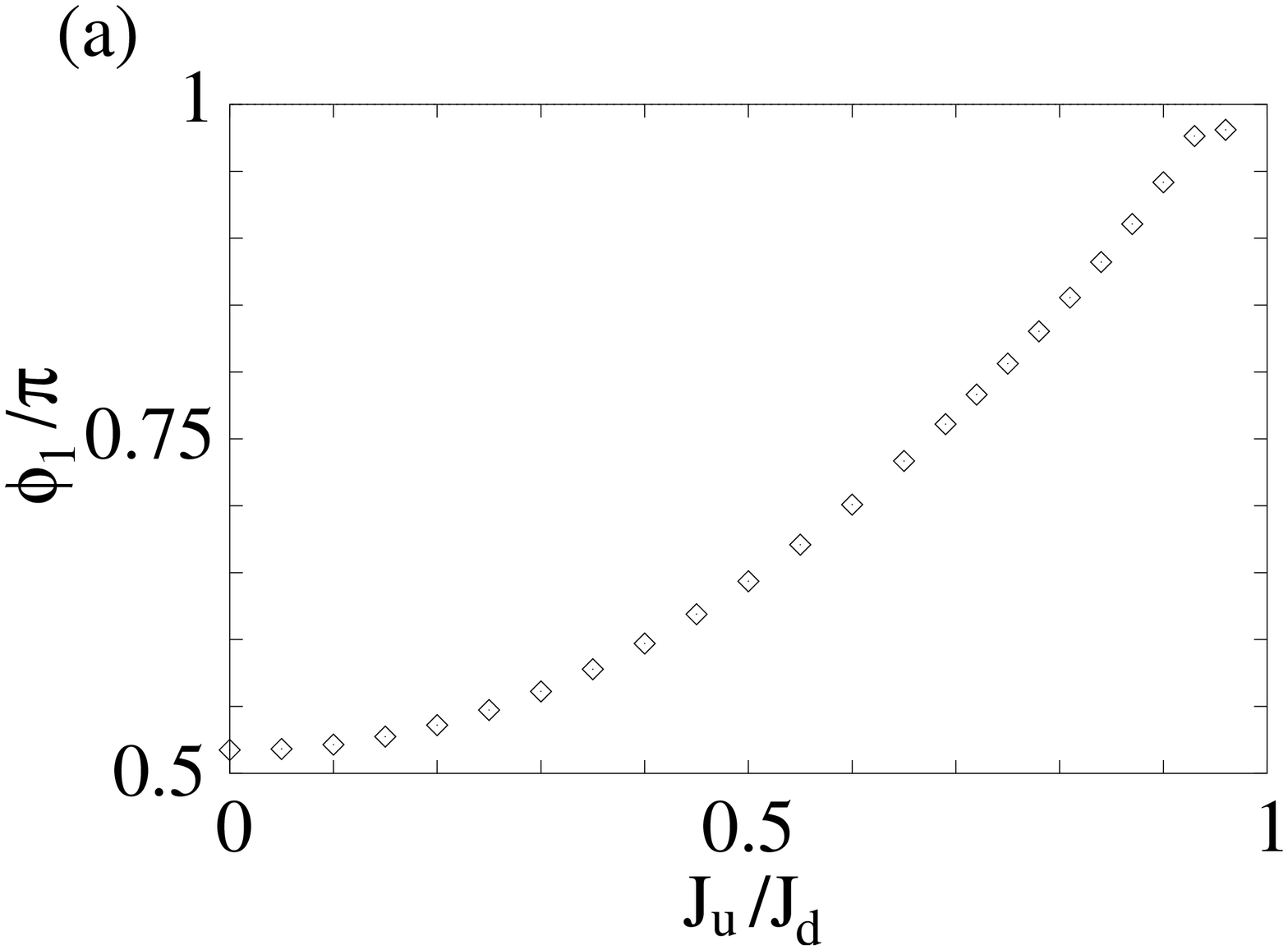}\includegraphics{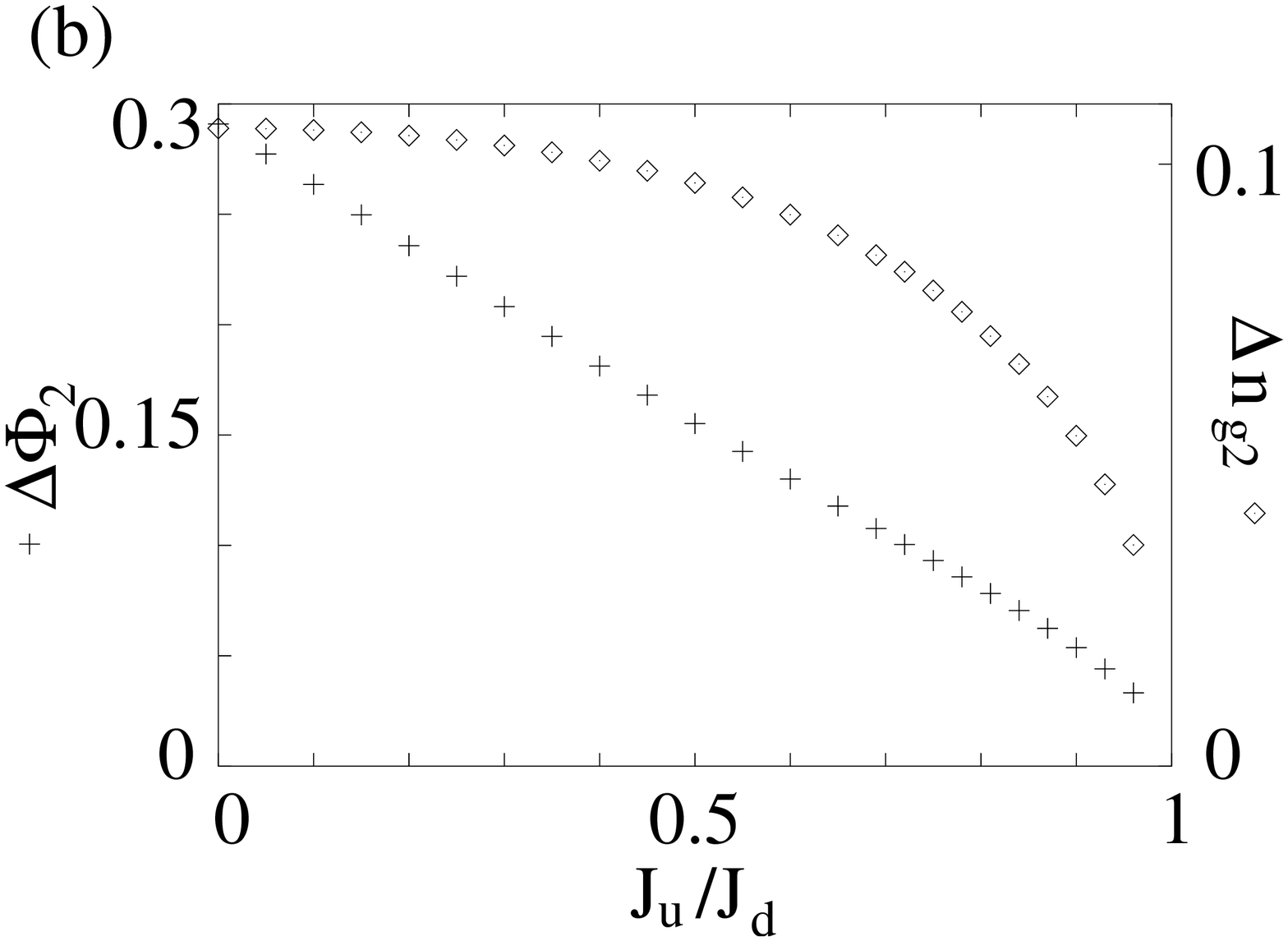}}}}
\caption{Dependence of the phase acquired during one turn around the
  region of enhanced field (a) and localization of the field
  (b) as functions of $J_u/J_d$.}
\label{phasedel}
\end{figure}

The phase shift acquired during one cycle can be controlled by the
strength of the Josephson couplings $J_u^{(1),(2)}$, which are tuned
by varying the  fluxes through the small SQUID loops. Let us
illustrate this effect for
the phase shift $\phi _1$. In Fig.\ref{phasedel}(a) we see the phase
acquired during one cycle as the function of $J_u/J_d$. The phase
shift increases as the asymmetry of the right SQUID is reduced,
and tends to the value $\pi$ (and for $J_u/J_d=1$ is exactly $\pi
$ as explained below). In Fig.\ref{phasedel}(b) we see that together
with reduction of the asymmetry of the SQUID, the width of the peaks
tends to zero. Thus the effect seems to be amplified together with enhanced
localization of the field.
Simple reasoning shows that for symmetric SQUIDs ($J_u = J_d$)
the effective field is zero, except for singular points. Moreover,
loops enclosing an odd number of these singularities 
correspond to the phase shifts $\phi _1 = \pm \pi$ and $\phi _2 = \pm
\pi$. Indeed, for symmetric SQUIDs the Hamiltonian in 
Eq.(\ref{eq:qua}) is real. The Berry field is, as seen from Eq.(\ref{eq:sua})
zero, unless there is a crossing in the 
energy difference between the ground state and the first excited
state. Such singularities are points of zero Josephson couplings ($\Phi
_1 = \Phi _2 = \Phi _m = 1/2$) and can be easily identified as the
triple points in the charging 
diagram (see Fig. \ref{honeyelect}). Depending on the transformation
we want to perform, one of 
the degenerate lowest states ($|\bar{0} \rangle $ or $|\bar{1} \rangle
$ and $|+ \rangle $ or $|- \rangle$) is for each singular point decoupled from
the first excited state, and the Berry phase will be zero for this
state, even if the path encloses the singularity. The complementary
state, however, is coupled [in the sense that the numerator in
Eq.(\ref{eq:sua}) is nonzero] to the excited state, and the phase shift
will be nontrivial. As only two states are involved in the
transformation, we can now use the spin-$1/2$ picture. The real
Hamiltonian $H = - B \cdot \sigma /2$  is parametrized by the $B$-field
which has only $x$ and $z$ nonvanishing components, so that each path
lies in the plane containing degeneracy. Cyclic 
evolution of the parameters corresponds to solid 
angles either $0$, when the loop does not
enclose the degeneracy, or $ \pm 2 \pi $, when it does. The corresponding
Berry phase is thus either $0$ or $ \pm \pi$ (see also
\cite{berry}). Applying the result to 
our system we see that the phases $\phi _1$ and $\phi _2$ can have
only values $0$ or $\pm \pi $, so that the resulting non-trivial
transformations within the subspace $\{|\bar{0} \rangle, |\bar{1} \rangle  \}$
are $\pm \sigma _z, \pm \sigma _x$. 

The latter transformation is a topological
charge pumping. This can be the easiest effect to observe experimentally as it
requires a simplified setup [see Fig. \ref{system}(b)], with symmetric
SQUID loops only, a single
voltage source, and a single current line generating equal control
fluxes $\Phi _1$ and $\Phi _2$ (assuming identical qubits). Let us
estimate the rate of the dynamical dephasing for this scheme.
Fluctuations of the fluxes $\Phi _1 = \Phi _2$, and of the gate
voltage $n_g$ do not violate the constraints that we put on the
parameters. Thus 
the most important source of errors is in this case the fluctuating flux $\Phi
_m$. Since we are interested in the degenerate subspace only, we apply
here the formula for the dephasing rate in the environment-dominated
regime, $\Gamma _{\phi } \approx 2 \pi \alpha k_B T /
\hbar$ \cite{mss}. Here $\alpha $ is the constant describing the
strength of the system-environment coupling. At the starting point,
where the dependence of $J_m$ on the flux is strongest, the
dephasing rate will be at maximum, and we take this value as our
estimate. At this point $\alpha = R_K (M J_{m}^0 \pi / \Phi _0^2)^2/4
R_I$, where the resistance of the external circuit inducing the flux,
$R_I$, is typically of the order of $100 \Omega $. $R_K \approx 25.8
\mbox{k}\Omega $ is the quantum resistance, and $M \approx 0.1 - 0.01$nH is
the mutual inductance between the SQUID and the control circuit. The
dephasing rate should be compared to the mean energy separation
between the excited and the ground state $\bar{\omega }_{e\rightarrow
  g}$, which constraints the time in which the loop is traversed. Then for 
$J_m^0 \approx 1$K and at $T = 20$mK we obtain $\Gamma _{\phi }/ \bar{\omega
}_{e\rightarrow   g} \approx 10^{-5} - 10^{-3}$. At optimal design
there should be thus 
much space left to both satisfy 
the adiabaticity condition and perform many charge pumping cycles
within the time $\Gamma _{\phi }^{-1}$.

To summarize, the results presented here prove the possibility to
perform robust 
geometric transformations in quantum systems by identifying points of
localized effective magnetic field. The resulting procedure resembles
the original setting of the Aharonov-Bohm effect. In this particular
system the geometric phases will be accompanied by dephasing
of dynamical origin, as the degeneracy of the ground state is controlled by
external parameters. However, the weak adiabaticity condition on
one hand and dominant role of the Berry field from the interior of the
enclosed region on the other make it possible to traverse the paths
quickly (and generate large phase shifts), before the dephasing
destroys the visibility of the effect.\\

The author thanks Yu.~Makhlin, R.~W.~Chhajlany, and R.~Fazio for
stimulating discussions and comments. This work was supported by the
DFG-Schwerpunktprogramm 
``Quanten-Informationsverarbeitung'', EU IST Project SQUBIT, and EC
Research Training Network.

\end{document}